\documentstyle[12pt]{article}

\begin{document}

\baselineskip=18pt
\null\vspace{2cm}
\begin{center}
{\bf CONSISTENT AND COVARIANT ANOMALIES\\
IN THE OVERLAP FORMULATION OF CHIRAL GAUGE THEORIES}\\
\vspace{2cm}
S. Randjbar--Daemi\\
International Centre for Theoretical Physics, Trieste, Italy.\\
and \\
J. Strathdee \\
35 Mana Street, Wellington 6002, New Zealand \\
\end{center}
\date{\today}
\vspace{1.5cm}
\centerline{ABSTRACT}
In this letter   we show how the covariant anomaly emerges in the
overlap scheme. We also prove that the overlap scheme correctly
reproduces the anomaly in the flavour currents such as $j^5_\mu$ in
vector like theories like QCD. \bigskip
\newpage

The overlap scheme is a framework for a lattice definition of chiral
gauge theories \cite{kn:NN}. It has been shown that it passes all
perturbative tests as long as no gauge fields are involved in the
internal loops \cite{kn:RS1}. In particular, the calculation of the
consistent chiral anomalies in the Yang-Mills \cite{kn:RS2} as well
as gravitational backgrounds \cite{kn:RS3} have been carried out and
shown to give the correct continuum limit. One aspect of anomaly
calculations, namely, the source and evaluation of the covariant
anomaly still needs some clarification. The distinction between
covariant and consistent anomalies was described and explained in
the pioneering work of Bardeen and
Zumino \cite{kn:BZ}. Here we consider this old question in the
context of lattice regularization. Our approach is to set up a
functional differential equation for the vacuum amplitude of
interest, the `chiral determinant'. This variational approach to the
lattice regularized amplitude is similar to the heat kernel method
adopted by Leutwyler \cite{kn:Leut} and we hope that it will be
helpful to make comparisons with Leutwyler's work as we go along.
Another question to be considered here concerns the anomalies in
global chiral symmetries such as the flavour singlet $U(1)$ axial
symmetry of QCD. This
must be examined if the overlap is going to be applied to a
lattice formulation of vector--like theories such as QCD.

In the overlap scheme one starts from two Hamiltonians, $ H_\pm (A)$,
 which differ from each other in the sign of a mass--like term. To obtain
chiral fermions in $D=2N$ dimensional Euclidean space the
Hamiltonians must describe the propagation of the fermions in a
$2N+1$ dimensional Minkowski space-time in the background of the
gauge field $A$. The 4-dimensional Euclidean space will be the plane
sitting at the origin of the 5th coordinate which is assumed to be
time-like \cite{kn:RS1}. Let $\vert A\pm\rangle$ denote the `Dirac' ground
states of
our Hamiltonians (negative energy fermion states are filled). They
satisfy the eigenvalue equations
\begin{equation}
H_\pm (A)\vert A\pm\rangle = E_\pm (A)\vert A\pm\rangle
\end{equation}
 We assume  $\vert A\pm\rangle$ are normalized $\langle A\pm\vert
A\pm\rangle =1$.
We shall give a general argument which is valid on lattice or
continuum but the explicit calculational checks  will be
carried out in the continuum only.

Under an arbitrary variation of $A$, eqn.(1) yields
\begin{equation}
(\delta E_+-\delta H_+)\vert A+\rangle +(E_+-H_+)\delta\vert A+\rangle =0
\end{equation}
whose general solution for $\delta \vert A+\rangle$ is
\begin{equation}
\delta\vert A+\rangle =G_+(\delta H_+-\delta E_+)\vert A+\rangle +i\
\Delta_+(\delta A,A)\vert A+\rangle
\end{equation}
where
$$
G_+(A)=\Bigl( 1-\vert A+\rangle\langle A+\vert\Bigr)\ {1\over E_+-H_+}\ \left(
1-\vert A+\rangle\langle A+\vert\right)
$$
Since $G_+\vert A+\rangle =0=\langle A+\vert G_+$ it follows that
 $\Delta_+$ can be obtained by multiplying (1) by $\langle A+\vert$,
\begin{equation}
i\ \Delta_+(\delta A,A)=\langle A+\vert\delta\vert A+\rangle
\end{equation}
It is easy to see that $\Delta (\delta A,A)$ is a real quantity.  Similar
formulae are valid for $\vert A-\rangle$.  We can therefore write the
variation of the overlap $\langle A+\vert A-\rangle$,
\begin{equation}
\delta\ \ell n \langle A+\vert A-\rangle = L(\delta A,A)+\ell (\delta A,A)
\end{equation}
where
\setcounter{equation}{0}
\renewcommand{\theequation}{5.\alph{equation}}
\begin{eqnarray}
L(\delta A,A) &=& {\langle A+\vert \delta H_+G_++G_-\delta
H_-\vert A-\rangle\over\langle A+\vert A-\rangle}\\[3mm]
\ell (\delta A,A)
&=& -i\Bigl(\Delta_+(\delta A,A)-\Delta_- (\delta A, A)\Bigr)
\end{eqnarray}
From these differential forms one can read out the overlap
expressions for the currents, $L_\mu$ and $\ell_\mu$ which we define to be
the coeficients of
$\delta A_\mu$ in $L$ and $\ell$ respectively.  It is clear that $L$,
and hence $L_\mu$, does not depend on the choice of the phases for
the states $\vert A\pm \rangle$.  (This is not true for $\ell$ and
$\ell_\mu$.)  Our claim is that $L_\mu$  reduces,
in the continuum limit, to the covariant current.  To show this we need to
examine
the response of various quantities to local gauge transformations.
\footnote{ Our $L$ and $\ell$ should be compared with the analogous
quantities in \cite{kn:Leut}}

Under a local gauge transformation $A\to A^\theta$, the Dirac ground states
$\vert A\pm\rangle$ undergo a transformation of the form
\setcounter{equation}{5}
\renewcommand{\theequation}{\arabic{equation}}
\begin{equation}
e^{iF_\theta}\vert A\pm\rangle =\vert A^\theta\pm\rangle e^{i\Phi_\pm
(\theta ,A)}
\end{equation}
where $\Phi_\pm$ are real angles and $F_\theta =\mathop{\sum}\limits_n\
\psi^\dagger (n)\ \theta (n)\ \psi (n)$ is an hermitian operator.
From (6) we draw several conclusions.

Firstly, define the effective action $\Gamma (A)$ by the overlap
formula
$$e^{-\Gamma (A)}={\langle A+\vert
A-\rangle\over\langle +\vert -\rangle}.$$  It follows from (6) that this
action functional transforms according to
\begin{equation}
\Gamma (A^\theta )-\Gamma (A)=i\Bigl(\Phi_+(\theta ,A)-\Phi_-(\theta ,A)\Bigr)
\end{equation}
This indicates that the real part of the effective action is gauge
invariant but its imaginary part may not be.  In fact, we have shown elsewhere
that the right hand side of (7) produces exactly the `consistent' anomaly
 \cite{kn:RS2}.

Next, assume $\theta$ is infinitesimal and expand (6) to first
order, obtaining
\begin{equation}
\delta_\theta\vert A\pm\rangle = i\ F_\theta\vert A\pm\rangle -i\ \Phi_\pm\vert
A\pm\rangle
\end{equation}
This implies
\begin{equation}
\langle A\pm\vert\delta_\theta\vert A\pm\rangle = i\langle A\pm\vert
F_\theta\vert A\pm\rangle -i\ \Phi_\pm
\end{equation}
or, using the definition (4) of $\Delta$, we can rewrite this in the form
\begin{equation}
\Delta_\pm (\delta_\theta A,A)=F_\pm (\theta ,A)-\Phi_\pm (\theta ,A)
\end{equation}
where  $F_\pm$ is defined by
\begin{equation}
F_\pm (\theta ,A)\equiv\langle A\pm\vert F_\theta\vert A\pm\rangle
\end{equation}
From  these equations we obtain
\begin{eqnarray}
\delta_\theta\Gamma (A) &=& i\Bigl(\Phi_+(\theta ,A)-\Phi_-(\theta
,A)\Bigr)\nonumber\\
&=& i\Bigl( F_+ (\theta ,A)-F_-(\theta ,A)\Bigr)
-i\Bigl(\Delta_+(\delta_\theta A,A)-\Delta_-(\delta_\theta A,A)\Bigr)
\end{eqnarray}

 We would like to show  that $i(F_+-F_-)$ is indeed the `covariant'
anomaly. The actual proof
of this assertion comes down to a direct calculation of $F_+-F_-$.
Therefore, we shall
evaluate $F_\pm$ up to first order terms in $\theta$ and second order terms
in $A$. This  will produce the leading part of the covariant anomaly in $D=4$
non--Abelian Yang--Mills theory.

First we remark that $F_\pm$, defined by the diagonal matrix elements (11),
are evidently covariant,
$$
F_\pm (\theta ',A')=F_\pm (\theta ,A)
$$
where
\begin{eqnarray}
\theta ' &=& g\theta g^{-1} \nonumber \\
A'       &=& gAg^{-1}+gdg^{-1} \nonumber
\end{eqnarray}
for any local gauge transformation, $g(n)$. This follows because the
transformation is realized by the action of a unitary operator,
$exp[iF_\omega]$, on the fermion field operators,
$$
e^{-iF_\omega}\psi (n)\ e^{iF_\omega}=g(n)\psi (n)
$$
where $F_\omega$ is an appropriate bilinear analogous to
$F_\theta$.  More generally, it is quite simple to see that
$L(\delta A^\theta,A^\theta)= L(\delta A,A)$  and hence that the current
$L_\mu$ is covariant.

To prove the above assertion we make a perturbative calculation in the
continuum. We start from the perturbative expression \cite{kn:RS2} for
$\vert A+\rangle$
\begin{equation}
\vert A+\rangle =\alpha_+(A)\Bigl[ 1-G_+(V-\Delta E_+)\Bigr]^{-1}\vert +\rangle
\end{equation}
where $G_+={1-\vert +\rangle\langle +\vert\over E_+-H_+}$ and $V=i\int
d^4x\ \psi^\dagger\ A\!\!\!/\ \gamma_5\ \psi$
Here $H_+=\int d^4x\ \psi^\dagger\ \gamma_5(\partial\!\!\!/\ +\Lambda )\psi$ and
$H_+\vert +\rangle =E_+\vert +\rangle$. $\alpha_+(A)$ is determined from the
normalization condition $\langle A+\vert A+\rangle =1$. We shall
adopt the Brillouin--Wigner phase convention according to which
$\alpha_+(A) >0$.
Since $V$ is first order in $A$ upto the desired order $F_+$ will be given by
\begin{eqnarray}
F_+ &=& \alpha^2_+\Bigl[\langle +\vert F_\theta\vert +\rangle +  1^{st}\
{\mathrm
order}\ +\Bigr.\nonumber\\
&&\Bigl.\langle +\vert F_\theta G_+VG_+V+VG_+F_\theta
G_+V+VG_+VG_+F_\theta\vert
+\rangle\Bigr]
\end{eqnarray}
In writing this expression we have assumed that the first order
contributions to
$\Delta E$ are zero (by Lorentz invariance in the continuum limit).

Define the second order terms $F_+^{(2)}$ by:
\begin{equation}
F^{(2)}_+=\langle +\vert
F_\theta G_+VG_+V+VG_+F_\theta G_+V+VG_+VG_+F_\theta\vert +\rangle
\end{equation}
It is easy to show that up to $A^2$ terms $\alpha_+$ is given by
$\alpha^2_+(A) =1-\langle +\vert VG^2_+V\vert +\rangle$.
We can then write the $A^2$ terms in (14) as
\begin{equation}
F_+=F^{(2)}_+-\langle +\vert VG^2_+V\vert +\rangle\langle +\vert F_\theta\vert
+\rangle
\end{equation}
 In all of the subsequent calculations we shall assume that the background
gauge field is
slowly varying in the scale of $\Lambda^{-1}$. We shall show  that the
second term in (16) will
not contribute to $F_+-F_-$. We shall thus
concentrate on the evaluation of $F^{(2)}_+$. This can be done by inserting
complete sets of states in (15). Here we shall sketch some of the calculational
steps.
After some relatively simple but lengthy calculations
\footnote{ The evaluation of the matrix elements have been discussed in more
detail in our earlier papers \cite{kn:RS2}.} we arrive at
\setcounter{equation}{0}
\renewcommand{\theequation}{17\alph{equation}}
\begin{eqnarray}
F^{(2)}_+ &=&
\int \left({dk\over 2\pi}\right)^4\ \left({dk'\over 2\pi}\right)^4\
\left({dp\over 2\pi}\right)^4\ {1\over (\omega_k+\omega_p
)(\omega_{k'}+\omega_p )}\nonumber\\[3mm]
&&\quad Tr\Bigl(\tilde\theta (p-k)U(k)\tilde A(k-k')U(k')\tilde A(k'-p)V(p
)\Bigr.\nonumber\\[3mm]
&&+\tilde A(p-k)U(k)\tilde\theta (k-k')U(k')\tilde A(k'-p)V(p)\nonumber\\[3mm]
&&+\tilde A(p-k')U(k')\tilde A(k'-k)U(k)\tilde\theta (k-p)V(p)\nonumber\\[3mm]
&&-\tilde\theta (k-p)U(p)\tilde A(p-k')V(k')\tilde A(k'-k)V(k)\nonumber \\[3mm]
&&-\tilde A(k-p)U(p)\tilde A(p-k')V(k')\tilde\theta (k'-k)V(k)\nonumber\\[3mm]
&&\Bigl.-\tilde A(k-k')V(k')\tilde A(k'-p)U(p)\tilde\theta
(p-k)V(k)\Bigr)
\end{eqnarray}
where $U(k) = [\omega _k+\gamma_5(ik\!\!\!/\ +\Lambda )]/2\omega _k=1-V(k)$ and
$\omega_k = (k^2+\Lambda^2 )^{1/2} $.
In evaluating the traces over the $\gamma$--matrices we need to consider only
the odd powers of $\Lambda$. (The even powers will cancel from $F_+-F_-$.)
After further lengthy calculations we arrive at a relatively simple
expression for the {\em pseudoscalar} part of $F^{(2)}_+$,
\begin{eqnarray}
F^{(2)}_+ &=& -\Lambda\int\left({dk\over 2\pi}\ {dk'\over 2\pi}\ {dp\over
2\pi}\right)^4\
{\varepsilon_{\mu\lambda\nu\sigma}\over\omega_k\omega_{k'}\omega_p}\ tr\Bigl(
\tilde\theta (p-k)\tilde A_\mu (k-k')\tilde A_\nu (k'-p)\Bigr)\nonumber\\[3mm]
&&\Biggl[ {k_\sigma p_\lambda + k'_\sigma k_\lambda\over
(\omega_k+\omega_p)(\omega_{k'}+\omega_p)}+
{k_\lambda k'_\sigma +k'_\lambda p_\sigma +k_\sigma p_\lambda\over
(\omega_p+\omega_{k'})(\omega_k+\omega_{k'})}\Biggr.\nonumber\\
&&\Biggl.\hspace{4.5cm} +{k_\lambda k'_\sigma +k_\sigma p_\lambda +k'_\lambda
p_\sigma\over (\omega_k+\omega_p)(\omega_{k'}
+\omega_k)}\Biggr]+\dots
\end{eqnarray}

We shall perform the integration over $p$ in the limit in which $A_\mu$ and
$\theta$ are slowly varying over distances of the order of $\Lambda^{-1}$. It
is more convenient to introduce $k_1$ and $k_2$ through $k= p+k_1+k_2$ and
$k' = p+k_2$ .
We can then write
\begin{equation}
F^{(2)}_+-F^{(2)}_- = \int\left({dk_1\over 2\pi}\ {dk_2\over 2\pi}\right)^4\
tr\left(\tilde\theta (-k_1-k_2)\ \tilde A_\mu (k_1)\tilde A_\nu (k_2)\right)\
G_{\mu\nu}(k_1,k_2)
\end{equation}
where in the limit of ${\vert k_i\vert\over\Lambda} \ll 1,\ \  i=1,2$;
\hskip .5cm
$G_{\mu\nu} (k_1,k_2)=
 -{\Lambda\over\vert\Lambda\vert}\ {1\over 8\pi^2}\
\varepsilon_{\mu\lambda\nu\sigma}k_{1\lambda}k_{2\sigma}$.
Thus
\begin{equation}
F^{(2)}_+-F^{(2)}_- ={1\over 8\pi^2}\ {\Lambda\over\vert\Lambda\vert}\int d^4x\
\varepsilon_{\mu\lambda\nu\sigma}\ tr\Bigl(\theta (x)\partial_\lambda
A_\mu\partial_\sigma A_\nu\Bigr)
\end{equation}
This is the correct leading order term for the covariant anomaly.
The second term of (16) actually vanishes for non--Abelian gauge
theories, because in this case $\langle +\vert F_\theta\vert
+\rangle =0$. This follows from the invariance of $F_+(\theta ,A=0)$
with respect to rigid gauge transformations. For $G=U(1)$ it can be
non--zero and divergent, in which case we should use a lattice
regularization. However, it will be independent of $\Lambda$. Thus
if non--zero $\langle +\vert F_\theta\vert +\rangle =\langle -\vert
F_\theta\vert -\rangle$. Therefore in the case $G=U(1)$ the
contribution of the second term in (16) to $F_+-F_-$ will be
$\langle +\vert VG^2_+V\vert +\rangle -\langle -\vert VG^2_-V\vert
-\rangle$. This quantity, being pseudoscalar and blinear in the
gauge field, must vanish in the continuum limit.  Hence, the second
term in (16) can be discarded in the Abelian case as well. This has been
verified by direct calculation.

Vector like theories such as QCD have no anomalies in the currents that
couple to vector potentials.  However, the axial vector current $j^5_\mu$
associated with the global chiral symmetry of massless quarks {\em is}
anomalous.  If this current is defined covariantly then its divergence is
given by

$$
\partial_\mu j^5_\mu ={1\over 16\pi^2}\ \varepsilon_{\mu\nu\lambda\sigma}\
tr\ F_{\lambda\mu}\ F_{\sigma\nu}
$$

It is easy to understand the derivation of this anomaly from the
formalism we have developed above.  To this end, assume that the
gauge group has the product structure, $G=G_{\mathrm local}\times
G_{\mathrm global}$.  In this case the infinitesimal gauge
transformation parameter will be the sum of two terms, $\theta
=\theta_{\mathrm local} +\theta_{\mathrm global}$, with
$\theta_{\mathrm local}$ and $\theta_{\mathrm global}$ lying in the
corresponding Lie algebras.  Likewise, the vector potential would be
written $A=A_{\mathrm local}+A_{\mathrm global}$ but with the
understanding that $A_{\mathrm global}$ will be set equal to zero,
and the infinitesimal parameter, $\theta_{\mathrm global}$ will be
constant (independent of $x$). Make these substitutions in
the formula (17.d), or rather, its fully covariantized generalization,
\setcounter{equation}{0}
\renewcommand{\theequation}{18}

\begin{eqnarray}
F_+-F_-&=&{1\over 32\pi^2}\ {\Lambda\over\vert\Lambda\vert}\int d^4x\
\varepsilon_{\mu\lambda\nu\sigma}\ tr\theta
F_{\lambda\mu}F_{\sigma\nu} \nonumber \\
       &=&{1\over 32\pi^2}\ {\Lambda\over\vert\Lambda\vert}\int d^4x\
\varepsilon_{\mu\lambda\nu\sigma}\ tr\Bigl( (\theta_{\mathrm
local}(x)+\theta_{\mathrm global}\ )F_{\mathrm local \ \lambda\mu}(x)
F_{\mathrm local \ \sigma\nu}(x)\Bigr)
\end{eqnarray}

The trace in these equations is understood to be taken over the
representation of $G_{\mathrm local}\times G_{\mathrm global}$ to which the
fermions belong. The coefficient of $\theta_{\mathrm global}$
defines the global anomaly. Clearly for a non Abelian
 $G_{\mathrm local}$ , only the $U(1)$ part of
$\theta_{\mathrm global}$ can survive in this formula. Because of
the assumed direct product structure, $G_{\mathrm local}\times
 G_{\mathrm global}$, the non--Abelian part of $G_{\mathrm global}$
cannot be anomalous. Of course, one could consider other covariant
currents, such as the axial vector colour octet in QCD which, by the
same logic, would be anomalous.

The curl of $\Delta_+-\Delta_-$ is another quantity which is of interst.
This quantity has been called $C$ in \cite{kn:Leut} and  is given by
\setcounter{equation}{0}
\renewcommand{\theequation}{19\alph{equation}}
\begin{equation}
C=\pm\ {1\over 2\pi}\int d^2x\ \varepsilon_{\mu\nu}\ tr(\delta_2 A_\mu\delta_1
A_\nu )
\end{equation}
in 2--dimensions and
\begin{equation}
C=\pm\ {1\over 8\pi^2}\int d^4x\ \varepsilon_{\mu\nu\lambda\sigma}\
tr\{\delta_1A_\mu ,\delta_2A_\nu\} \partial_\lambda A_\sigma
\end{equation}
in 4--dimensions.

Our starting point will be the perturbative expansion (13). To obtain the above
results we need to go up to $A^3$ terms and substitute the result in the
definition of
$\Delta_+(\delta A,A)$.
 This series can be used to evaluate the curl of
$\Delta_+$. In the first step we get
\setcounter{equation}{0}
\renewcommand{\theequation}{20}
\begin{eqnarray}
&&i\Bigl(\delta_2\Delta_+(\delta_1A,A)-
\delta_1\Delta_+(\delta_2A,A)\Bigr) =\nonumber\\
&&\quad = 2\delta_2\alpha\langle A+\vert\delta_1(GV)+\delta_1(GV)^2+\dots \vert
+\rangle\nonumber\\
&&\qquad +\alpha^2\langle
+\vert\Bigl(\delta_2(VG)+\delta_2(VG)^2+\dots \Bigr) \Bigl( \delta_1(GV)+
\delta_1(GV)^2+\dots\Bigr) \vert +\rangle\nonumber\\
&&\hspace{8cm} -(1\leftrightarrow 2)
\end{eqnarray}
We have dropped the energy shifts $\Delta E$. If present they will
contribute disconnected pieces. Also $\alpha (A)$  contributes
disconnected pieces, if it contribute at all to the curl
$(\Delta_+-\Delta_-)$.
It is not difficult to see that in $D=2$ and $D=4$ the $\delta\alpha$ terms do
not contribute to the curl of $\Delta_+$.
\footnote{We have shown in \cite{kn:RS1} that the disconnected contributions
take care of themselves and they do not contribute to the effective
action.}  Equation (20) then reduces to

\begin{eqnarray*}
i\ curl\ \Delta_+ &=& i\Bigl(\delta_2\Delta_+(\delta_1A,A)
-\delta_1\Delta_+(\delta_2A,A)\Bigr)\\
&=&\langle
+\vert\left(\delta_2(VG)+\delta_2(VG)^2+...\right)
\left(\delta_1(GV)+\delta_1(GV)^2+...\right)\vert
+\rangle\\
&&\hspace{8cm} -(1\leftrightarrow 2)\\
&=& C^{(0)}_++C^{(1)}_++C^{(2)}_++\dots
\end{eqnarray*}
where
\setcounter{equation}{0}
\renewcommand{\theequation}{21.\alph{equation}}
\begin{eqnarray}
C^{(0)}_+ &=& \langle +\vert\delta_2(VG)\delta_1(VG)\vert +\rangle
-(1\leftrightarrow 2)\\[3mm]
C^{(1)}_+ &=& \langle
+\vert\delta_2(VG)\delta_1(GV)^2+\delta_2(VG)^2\delta_1(GV)\vert +\rangle -
(1\leftrightarrow 2)\\
\vdots\ \ \  &&\nonumber
\end{eqnarray}
Our aim is to identify the $\Lambda$--odd terms in $C^{(i)}_+$, for slowly
varying vector potentials $A_\mu (x)$. The $\Lambda$--even terms will be
cancelled from $C^{(i)}_+-C^{(i)}_-$. We shall show that $C^{(0)}_+$ will
contribute only in $D=2$ and its contribution will correctly reproduce (19.a).
Likewise $C^{(1)}_+$ will contribute in $D=4$ and its contribution will
correctly reproduce (19.b).\\

\noindent\underline{$C^{(0)}_+$}:  It is not hard to show that
\begin{eqnarray*}
C^{(0)}_+ &=& -\int\left({dk_1\over 2\pi}\ {dk_2\over 2\pi}\right)^D\ {1\over
(\omega_{k_1}+\omega_{k_2})^2}\ \Bigl[ tr\left(\delta_2\tilde A_\mu
(k_2-k_1)\delta_1\tilde A_\nu(k_1-k_2)\right)\Bigr.\\
&&\hspace{3cm} \Bigl. \cdot tr\Bigl(\gamma_5\gamma_\mu
U(k_1)\gamma_5\gamma_\nu
V(k_2)\Bigr) -\delta_1\leftrightarrow\delta_2\Bigr]
\end{eqnarray*}
The $\gamma$--traces  will vanish for $D>2$. For $D=2$ they can easily be
evaluated and
we obtain
\begin{eqnarray*}
C^{(0)}_+ &=&-i\ \varepsilon_{\mu\nu}\ \Lambda\int\left({dp\over
2\pi}\right)^2\ tr\ \delta_2\tilde A_\mu (-p) G(p)\delta_1\tilde A_\nu(p)\\
&&\hspace{5cm} +\Lambda\ {\mathrm even\ terms}
\end{eqnarray*}
where
\begin{eqnarray*}
G(p) &=&\int \left({dk\over 2\pi}\right)^2\ {1\over\omega (k+{p\over 2})\omega
(k-{p\over 2})\left(\omega (k+{p\over 2})+\omega (k-{p\over 2})\right)}\\
&=&\ {1\over
2}\int\left({dk\over 2\pi}\right)^2\ {1\over\omega^3(k)} ={1\over 4\pi}\cdot
{1\over\vert\Lambda\vert}
\end{eqnarray*}
where the integral over p has been calculated in the limit of
${\vert p\vert\over\Lambda}\mathop{\to 0}$.
Thus
$$
C^{(0)}_+=-i\ {\Lambda\over\vert\Lambda\vert}\ {1\over 4\pi}\int d^2x\
\varepsilon_{\mu\nu}\ tr\ \delta_2A_\mu(x)\ \delta_1A_\nu(x)
$$
and

Therefore $\delta_2\left(\Delta_+(\delta_1A,A)-\Delta_-(\delta_1A,A)\right)
-(1\leftrightarrow 2) =C$
where $C$ is given by (19.a).\\

\noindent\underline{$C^{(1)}_+$}:  The evaluation of $C^{(1)}_+$ is
considerably more lengthy. Firstly it can be
shown that as $\Lambda\to\infty$, in $D=2, C^{(1)}_+=0$. Secondly the
contribution of the matrix elements of $\delta_2(VG)\delta_1(GV)^2$ in
(21.b) in D=4
are given by
\begin{eqnarray*}
&&\langle +\vert\delta_2(VG)\delta_1(GV)^2\vert +\rangle =\\
&&\qquad =\langle +\vert \left(\delta_2V G^2 \delta_1
VGV+\delta_2VG^2VG\delta_1V\right)\vert +\rangle\\[3mm]
&&\langle +\vert\delta_2VG^2\delta_1VGV\vert +\rangle -(1\leftrightarrow 2) =\\
&&\qquad =-{\Lambda\over\vert\Lambda\vert}\ {i\over 96\pi^2}\
\varepsilon_{\mu\nu\lambda\sigma}\int d^4x\ tr\{\delta_2A_\mu
,\delta_1A_\nu\}\partial_\sigma A_\lambda\\[3mm]
&&\langle +\vert\delta_2VG^2VG\delta_1V\vert +\rangle - (1\leftrightarrow 2) =\\
&&\qquad ={\Lambda\over\vert\Lambda\vert}\ {2i\over 96\pi^2}\
\varepsilon_{\mu\nu\lambda\sigma} \int d^4x\ tr\Bigl\{
(\delta_1A_\lambda\delta_2A_\mu +\delta_2A_\mu
\delta_1A_\lambda)\partial_\sigma A_\lambda\Bigr\}
\end{eqnarray*}
The contribution of the matrix element of $\delta_2(VG)^2\delta_1(GV)$ in
(21.b) can be obtained from that of $\delta_2(VG)\delta_1(GV)^2$ by complex
conjugation followed by an interchange of $1\leftrightarrow 2$. We can then
assemble all these results to obtain
$$
C^{(1)}_+={\Lambda\over\vert\Lambda\vert}\ {i\over 96\pi}\
\varepsilon_{\mu\nu\lambda\sigma} \int d^4x\
tr\Bigl[ (-2-4)\{\delta_2A_\mu ,\delta_1A_\nu\}\partial_\sigma A_\lambda\Bigr]
$$
Therefore
$$
C^{(1)}_+-C^{(1)}_-=-{\Lambda\over\vert\Lambda\vert}\ {i\over 8\pi}\
\varepsilon_{\mu\nu\lambda\sigma} \int d^4x\
tr\{\delta_2A_\mu ,\delta_1A_\nu\}\partial_\sigma A_\lambda
$$
This is in agreement with the expression (19.b) given by Leutwyler
\cite{kn:Leut}.\\

\noindent{\bf Acknowledgments}
 We are appreciative of communications with Rajamani Narayanan and
Herbert Neuberger.

\end{document}